# The effects of coating culture dishes with collagen on fibroblast cell shape and swirling pattern formation

Kei Hashimoto[1,2,3,†], Kimiko Yamashita[1,2,4,5,6,†], Kanako Enoyoshi[1,2,†], Xavier Dahan[2,7], Tatsu Takeuchi[8], Hiroshi Kori[1,9,*], and Mari Gotoh[3]

[1]Graduate School of Humanities and Sciences, Ochanomizu University, Ohtsuka, Bunkyo-ku, Tokyo, Japan

[2]Program for Leading Graduate Schools, Ochanomizu University, Ohtsuka, Bunkyo-ku, Tokyo, Japan

[3]Institute for Human Life Innovation, Ochanomizu University, Ohtsuka, Bunkyo-ku, Tokyo, Japan

[4]Physics Division, National Center for Theoretical Sciences, Hsinchu, Taiwan

[5]Department of Physics, National Tsing Hua University, Hsinchu, Taiwan

[6]Institute of High Energy Physics, Chinese Academy of Sciences, Beijing 100049, China

[7]Institute for Excellence in Higher Education, Tohoku University, Sendai, Japan

[8]Department of Physics, Virginia Tech, Blacksburg, VA 24061, USA

[9]Department of Complexity Science and Engineering, Graduate School of Frontier Sciences, The University of Tokyo, Kashiwa, Japan

*Correspondence should be addressed to H. Kori (kori@k.u-tokyo.ac.jp)

# Abstract

Motile human-skin fibroblasts form macroscopic swirling patterns when grown to confluence on a culture dish. In this paper, we investigate the effect of coating the culture-dish surface with collagen on the resulting pattern, using human-skin fibroblast NB1RGB cells as the model system. The presence of the collagen coating is expected to enhance the adherence of the fibroblasts to the dish surface, and thereby also enhance the traction that the fibroblasts have as they move. We find that, contrary to our initial expectation, the coating does not significantly affect the motility of the fibroblasts. Their eventual number density at confluence is also unaffected. However, the coherence length of cell orientation in the swirling pattern is diminished. We also find that the fibroblasts cultured in collagen-coated dishes are rounder in shape and shorter in perimeter, compared to those cultured in uncoated polystyrene or glass culture dishes. We hypothesize that the rounder cell-shape which weakens the cell–cell nematic contact interaction is responsible for the change in coherence length. A simple mathematical model of the migrating fibroblasts is constructed, which demonstrates that constant motility with weaker nematic interaction strength does indeed lead to the shortening of the coherence length.

# 1. Introduction

Collective cell migration is a key process observed at various stages in the development of multi-cellular organisms, starting with gastrulation and continuing into organogenesis [1]. Well-studied examples include neural-tube closure of vertebrae, and lateral-line formation in zebrafish. After birth, it is involved in wound healing and cancer metastasis [2]. Collective cell migration is also observed in single-cell organisms. A well-known example is aggregation by which Dictyostelium cells form a slug-like structure when starved [3]. Deciphering the mechanisms that drive robust and precise collective migration of large numbers of cells is of vital importance in understanding development, differentiation, and evolution, with many possible applications in cancer therapies, regenerative medicine, and tissue engineering [2, 4-6].

*In vitro* cultivation studies can provide important insights into collective cell migration. When cultivated densely, complex alignment patterns are known to spontaneously form in several types of cells. The characteristic coherence length of the resulting alignment pattern is of special interest since it provides a measure of the number of cells that can migrate collectively. One of the key factors determining the coherence length is the strength of the cell–cell contact interaction. Although a migrating cell has vectorial polarity, moving toward a specific heading, the alignment of these cells is often nematic, that is, neighbouring cells tend to migrate either in parallel or anti-parallel directions. This nematic alignment can be observed in several cellular-scale objects including gliding microtubules, actin filaments, bacteria, and cultured cells [7-14]. Nematic alignment is also observed in purely mechanical systems whose members interact via the excluded-volume effect, such as in a population of rod-like objects [15]. In this case, the strength of the nematic interaction is determined by the shape of the rods; the interaction being stronger between longer rods.

In the present paper, we investigate the alignment pattern of human-skin fibroblast NB1RGB cells grown to confluence in a culture dish. Skin-fibroblasts are cells that provide structural support for the skin and are easily cultivated *in vitro*. They represent a convenient model system for studying the collective migration of cells. Here, we examine how the alignment pattern of the fibroblasts depends on whether or not the dish surface is coated with collagen. To check for possible dependence of the result on the substrate material, the experiment is performed on two type of dishes: polystyrene and glass.

Since fibroblasts adhere to collagen via integrins, a natural expectation would be that the collagen-coated surface would provide enhanced traction to the fibroblasts leading to their enhanced motility, which in turn would lead to a change in the alignment pattern. *In vitro*, the collagenous extracellular matrix (ECM) is known to stimulate skin-fibroblast motility [16], which adds support to this expectation.

As reported in Ref [12], when fibroblasts are seeded onto a culture dish at low density, the cells adhere to the dish individually and then migrate randomly into cell-free areas while only occasionally coming into contact with other cells. As they proliferate, their density increases and confluence is eventually achieved. Confluent fibroblasts align locally along their elongated axes and form macroscopic swirling patterns. (See Supplementary Videos 1 and 2.) A similar swirling pattern, the storiform pattern, is often observed in fibrohistiocytic lesions *in vivo* [17, 18].

We find that for both polystyrene and glass dishes, the characteristic coherence length of the swirling pattern decreases as the density of the collagen coated onto the dish surface is increased. Moreover, we observe that the cells become rounder in shape with the increase in coated-collagen density, whereas cell-number density and, unexpectedly, cell motility remain unchanged. From these experimental results, we hypothesize that the difference in the coherence length mainly follows from the difference in the strength of the nematic contact interaction between the cells; i.e., rounder cells experience weaker nematic interactions, and thereby the coherence length becomes shorter. To test the feasibility of this hypothesis, we construct a simple mathematical model of migrating cells in which the strength of the nematic interaction can be controlled by a single parameter. Numerical simulations of this model demonstrate that the coherence length does indeed correlate positively with the nematic interaction strength. We thus propose that the collagen coating first leads to the change in the fibroblast cell-shape, which in turn shortens the coherence length. Possible mechanisms in which the collagen coating leads to the rounding of the fibroblasts is also discussed.

## 2. Results

**Human-skin fibroblast NB1RGB cells form macroscopic swirling patterns**

Human-skin fibroblast NB1RGB cells form macroscopic swirling patterns when cultivated in a culture dish due to their elongated shape, motility, and proliferation within the confined two-dimensional surface. Our objective is to quantify the difference in the swirling patterns when the fibroblasts are cultured on uncoated, and collagen-type-I-coated dishes. As discussed above, we perform the experiment on both polystyrene- and glass-bottom dishes. The collagen-coating procedure is detailed in the Materials and Methods section. The resulting collagen density on the surface of the culture dish depends on the concentration of collagen type-I in the initial collagen solution used in the coating process. We use the coating obtained from a 10.0 μg/mL collagen type-I solution as the standard reference.

Typical results for uncoated and collagen-coated dishes are shown in Fig. 1. See also Supplementary Videos 1 (uncoated) and 2 (collagen-coated). In the first row of Fig. 1A, still images of the confluent fibroblast swirling patterns on uncoated (left) and collagen-coated (right) polystyrene dishes are shown. The black scale bar in the lower-right corner of the right image is 1 mm long. For ease of comparison, the orientations of the individual fibroblasts in these images are read using Orientation J software [19] and shown colour-coded in the second row of Fig. 1A, with light-blue and red respectively indicating horizontal and vertical orientations (See "Orientation Analysis" in Materials and Methods for details). Other colours indicate orientations in between, as shown on the scale on the right margin of Fig. 1B.

Comparing the colour-coded images by inspection, one discerns that the fibroblasts form into patches of cells with similar orientation, and that these patches are slightly larger for the uncoated dish compared to the collagen-coated dish.

We quantify this observation by extracting the correlations of cell orientations from the images following the procedure detailed in Materials and Methods. First, the 1600 × 1200 pixel image is divided into a 50 × 38 grid, each subdivision being 32 × 32 pixels in size. Then, the block-averaged orientation is calculated for each subdivision, the results of which are shown in Fig. 1B. The correlations between the block-averaged orientations for each separation of the blocks is then calculated.

The resulting correlation functions are plotted in Fig. 1C (uncoated polystyrene) and 1D (collagen-coated polystyrene) for the images shown in Fig. 1A. A blow-up of the graphs between 0 to 1 mm is shown in the subframe inside Fig. 1D (labelled C' and D'). We can see that the orientation correlation falls off more quickly for the collagen-coated dish compared to the uncoated dish. This difference is consistently reproduced for multiple dishes, over multiple repetitions of the experiment, for both dish materials.

To characterise the fall-off of the correlation function with distance, we define the length $d_{50}$ to be the distance in which the correlation is reduced to one-half (50%) of the maximum value (100%). How this length is determined is illustrated in Figs. 1C and 1D. The values of $d_{50}$ are calculated following this procedure for multiple images and averaged. 40-image averages of $d_{50}$ for polystyrene are plotted in Fig. 1E for the uncoated, and three collagen-coated cases, where initial collagen solutions of concentrations 0.1, 1.0 and 10.0 $\mu g/mL$ are respectively used. Fig. S3B in Supplementary Information compares 4-image averages of $d_{50}$ for uncoated and coated-from-10.0 $\mu g/mL$-collagen glass dishes.

A clear trend can be seen in Fig. 1E in which the orientation coherence length decreases with the initial collagen concentration, indicating that the cell orientations vary more rapidly with distance when the collagen coatings on the culture dishes are denser, resulting in the cells forming more swirling patterns overall. Fig. S3B, though having only two data points, confirms this trend.

The culture dishes in this study are coated with collagen molecules dissolved in acidic conditions, and not with collagen fibrils. To test whether the structure of the collagen coating affects the coherence length, we culture the cells in dishes coated with various types of collagen, all starting from an acidic (pH 3.0) solution of concentration 10.0 $\mu$g/mL. These are:

1. Collagen type-I (the reference)
2. Heat-denatured collagen type-I
3. Gelatin, which comprise thermal denatured collagen fibrils,
4. Collagen type-IV, which is a non-fibrillar form of collagen.

In all of these cases, the values of $d_{50}$ are decreased compared with those of the control (uncoated). See Figs. S2A through F in Supplementary Information. This indicates that the reduction of orientation coherence occurs independently of the structure of collagen.

**Collagen coating affects cell shape, but not single-cell motility or cell-number density**

To clarify the effects of the collagen coating on individual fibroblast cells, we analyse cell motility and cell-number density, both of which are likely to affect the alignment pattern [20, 21]. To investigate cell motility, we track the movements of individual cells cultured at low and high densities (Fig. 2A). This measurement is possible for glass dishes only due to limitations of our equipment. At low density, the instantaneous velocity is slightly smaller for the collagen-coated dish compared to the uncoated dish, but not by a significant amount. At high cell density, the instantaneous velocities are indistinguishable between the uncoated and collagen-coated cases. A similar tendency is observed in the 10h-average velocity (Figs. S3D, E), suggesting that directional persistency in the cell motility has little dependence, if any, on the dish coating. Moreover, we measure the number of cells in the high-density cultures, for both dish materials, and find no significant difference in the numbers of cells (Figs. 2B and S3C).

We also analyse the effects of the collagen coating on the fibroblast cell shape. We measure the area $S$ and perimeter $L$ of each cell on uncoated and 0.1, 1.0, and 10.0 μg/mL collagen type-I-coated polystyrene dishes at low density (Figs. 2C and 2F), and quantify the cell roundness in terms of the circularity $4\pi S/L^2$ (Fig. 2E). The comparison between uncoated and 10.0 μg/mL collagen type-I-coated glass dishes is shown in Figs. S3F, G in Supplementary Information.

These measurements indicate that the NB1RGB fibroblasts at low cell density become rounder and smaller as the density of the collagen on the coating is increased, for both polystyrene and glass dishes. The same tendency is observed for heat-denatured collagen type-I, gelatin, and collagen type-IV on polystyrene (Figs. S2G, H), which suggests that interactions with the collagen coating makes the shape of NB1RGB fibroblasts rounder and smaller. At high cell density, it is difficult to quantify the cell shape since the cell boundaries become difficult to discern (Fig. 2D). Regarding cell area, there should be no significant difference between the uncoated and 10.0 μg/mL collagen type-I-coated dishes at high cell density given that the final cell densities are indistinguishable.

**Simulation: interaction strength increases coherence length**

Among our observations, the only significant difference between the cells cultured in coated and uncoated dishes is in their shapes at low density. Although it is possible that the difference in the cell shape is not maintained at high density, the shape of isolated single cells is likely to affect the orientation dynamics in highly dense cell populations.

Motivated by previous studies on the collective dynamics of rod-like objects [15], we consider the hypothesis that the nematic interactions between the cells in collagen-coated dishes are weaker than those in the uncoated dishes due to the cells becoming rounder in the presence of collagen, and it is this weaker nematic interaction that leads to the shorter coherence length.

The viability of this hypothesis depends on whether varying the strength of nematic interactions alone is sufficient in changing the coherence length. To this end, we construct a simple mathematical model with just a few tunable parameters, as described in Materials and Methods. The model includes effects of cell motility, cell proliferation, excluded volume effect, and nematic interactions. We run numerical simulations of our model to assess the effect of different nematic interaction strengths on the cell alignment patterns.

Despite its simplicity, our model reproduces the experimentally observed dynamical behaviour well. See Supplementary Videos 3 and 4. Fig. 3 shows some typical numerical results, where the parameter $K$ (in units of hour$^{-1}$) quantifies the strength of the nematic interaction. Fig. 3A is the alignment pattern obtained for $K = 0.020/h$, after running the simulation from a random initial condition for an equivalent of 144h. Fig. 3C is its block-averaged cell orientations, and Fig. 3E the corresponding distance dependence of the orientation correlation. Figures 3B, 3D, and 3F are those for $K = 0.015/h$.

As shown in Fig. 3G, an increase in the nematic interaction strength $K$, with all other model parameters kept fixed, results in an increase in the coherence length. Fig. 3G also shows how the $K$-dependence of $d_{50}$ changes with the cell migration velocity $v$. The three values of $v$ chosen are the experimentally observed values at low cell-density with and without collagen coating (Fig. 2A) and their average. There is little variation demonstrating the independence of our conclusion to $v$. We also note that the observed values of $d_{50}$ develop rather late in the simulation as shown in Fig. 3H.

To gain deeper understanding of the pattern formation process, we perform further numerical simulations under various conditions. The results are shown in Fig. 4 and Supplementary Videos 5 - 7. In all of the cases except for $K = 0$, we confirm the formation of swirling patterns and the increase of $d_{50}$ values with the strength of nematic interaction.

Recall that in Fig. 3H, $d_{50}$ values grow rapidly at late times. In contrast, in the "from high density" case in Fig. 4B, $d_{50}$ grows from earlier times. These observations suggest that the growth of the coherence length starts after the cell density becomes sufficiently high.

To check the robustness of our results, we further perform additional numerical simulations in which noise is included in the angular dynamics (Figs. 4C, D and Supplemental movies 6, 7). We introduce white noise of the strength $\mu$ (with units $\text{hour}^{-1}$) in the angular dynamics of each cell. The correlation time of angular dynamics of single isolated cells is approximately given by $1/\mu$, i.e., the direction of a cell at time $t$ is approximately independent of that at time $t - 1/\mu$. We choose a reference value of $\mu$ to be 0.1/h in our simulations. As seen in the movies, swirling patterns develop even in the presence of noise though the strength of nematic interaction required is larger compared to the noiseless case. In Fig. 4C, it is observed that substantial coherence emerges when the coupling strength $K$ is comparable to or larger than the noise strength $\mu$=0.1/h, and the coherence length increases with the coupling strength $K$. Conversely, Fig. 4D shows the dependence of $d_{50}$ on the noise strength $\mu$ with $K$ fixed to 0.20/h, and the value of $d_{50}$ can be seen to increase as the noise strength $\mu$ is lowered. As there is no discernible experimental difference in the directional persistency between coated and uncoated dishes (Figs. S3D, E), the difference in actual noise level must be small, and it is unlikely such a difference would account for the observed difference between the $d_{50}$ values in the fibroblast cultures.

Our simple mathematical model thus demonstrates that the effect of collagen coating on the orientation patterns can be accounted for solely by collagen's effect on the strength of the nematic interactions between the fibroblast cells.

## 3. Discussion

Human-skin fibroblasts cultured at high density form macroscopic swirling patterns in the culture dishes. This study reveals that the cells become rounder and form less coherent patterns as the density of the collagen coated on the culture dishes is increased.

There are clear differences in the morphological properties between cells cultured in the uncoated and coated dishes; cells cultured in the coated dishes are shorter in perimeter and rounder (Figs. 2C, E, F). The following molecular mechanism could underlie this change. It is known that the elongated cells have developed stress fibres; conversely, lamellipodia, which is an actin projection on the edge of the cell, rather than a stress fibre, is activated to make the cells round [22]. Two members of the Rho GTPase family, RhoA and Rac1, are necessary for the regulation of various cellular behaviours, including microfilament network organisation. The formation of stress fibre and lamellipodia are induced by RhoA and Rac1, respectively [23-27]. RhoA and Rac1 inhibit each other's activation, and this competition between the two is one of the factors that regulate cell morphogenesis [23, 28]. It has been reported that collagen type-I increases the activity of Rac1 in human platelets [29]. In fibroblasts, it is unknown whether collagen controls the Rac1 activity, whereas the inhibition of Rac1 activation promotes the expression of collagen protein, suggesting that collagen interacts with Rac1 in fibroblasts [30]. Therefore, it is possible that the collagen coating induces the development of lamellipodia and decreases that of stress fibres *via* Rac1 activation, thus changing the NB1RGB cell roundness.

Collagen is also known to control fibroblast motility. Reference [16] reports that collagen coating promotes cell migration into cell-free space in scratch assays. This result seemingly differs from our results shown in Fig. 2A, wherein no significant difference was found between the migration speeds in the uncoated and coated dishes. However, they do not necessarily contradict each other since Ref. [16] studied a different scenario: it analysed the migration of the cell assembly from densely populated to scratched regions. Moreover, the migration speed of individual cells may be sensitive to its measurement method, such as the discretisation of cell trajectories and the observation time-interval. Depending on the method, a significant difference between the uncoated and coated dishes could arise. However, our numerical simulations reveal that the coherence length is insensitive to small variations in migration velocity (Fig. 3G), suggesting that the difference in migration velocities at low density, if any actually exists, does not have a substantial effect on pattern formation at high density.

As the collagen density is increased, the cells become rounder, whereas cell motility remains unchanged, as shown in Figs. 1E and 2E. The same tendency is found for different types of coating materials (collagen type-I, heat-denatured collagen type-I, gelatin, and collagen type-IV), as shown in Fig. S2G. We thus hypothesise that the change in coherence length results from a change in the cell–cell contact interactions mediated by a change in cell shape. However, we do not have any direct evidence of this, and it therefore remains an open issue. Use of genetic manipulation or inhibitors that exclusively control the cell shape would be required for such a study.

Concerning the mathematical models of fibroblast orientation proposed in the literature, some focus on the interplay between the cells and the extra-cellular matrix (ECM) [31, 32], known as dynamic reciprocity. Fibroblast orientation has also been modelled in terms of individual cell migration [33]. Systems of reaction-diffusion and integro-partial differential equations have also been used to model fibroblast orientation; however, these require large numbers of parameters and computational complexity that would unnecessarily complicate the isolation of the key target parameter. Instead, our goal in using mathematical modelling is to test the hypothesis that changes in the coherence length of the orientation patterns can be driven solely by changes in the strength of cell-cell interactions. We therefore consider a model in which the strength of the nematic interaction can be controlled by a single parameter. In addition, to keep the model as simple as possible, we regarded each cell as a self-driven particle with constant speed, following various theoretical studies on collective migration [34, 35]. Despite the simplicity of our model, it qualitatively reproduces our experimental findings (Fig. 3) and represented rather realistic dynamics of collective migration (Supplementary Videos 3 and 4). Our model thus demonstrates that the coherence length increases with the strength of the cell–cell interaction, lending support to our hypothesis.

Using our mathematical model, we perform various *in silico* experiments (Fig. 4, Supplementary Videos 5,6). Similar swirling patterns are observed in all the conditions we employ except for the case of $K = 0$. In particular, qualitatively the same results are obtained even when spontaneous mobility, reproduction, and excluded volume effect are all turned off. Therefore, we propose that the nematic interaction is the primal factor for the alignment process and the formation of swirling patterns in cell cultures.

When only the nematic interaction is considered, the system is essentially the same as a population of identical oscillators distributed in a two-dimensional space with a synchronization interaction between close neighbours. The synchronization process of such a system can be described by the model $\frac{d\theta_i}{dt} = \omega + \widehat{K} \sum_{j=\langle i \rangle} \sin m(\theta_j - \theta_i)$ [36], where $\omega$ denotes the natural frequency of each oscillator, $\widehat{K}$ denotes the interaction strength, and $j = \langle i \rangle$ indicates a sum over the nearest neighbours of the $i$-th oscillator. The parameter $m$ is introduced for convenience to toggle between the two cases of ferromagnetic ($m = 1$) and nematic ($m = 2$) interactions. The latter corresponds to our situation, whereas the former is usually considered for a system of interacting oscillators. We may set $\omega = 0$ without loss of generality as it corresponds to the change $\theta_i \to \theta_i - \omega t$. The parameter $m$ may also be set to unity without loss of generality as it corresponds to the changes $m\theta_i \to \theta_i, m\omega \to \omega, m\widehat{K} \to \widehat{K}$. Thus, let us assume $\omega = 0$ and $m = 1$. Then, it is clear that $\widehat{K}$ determines only the time scale of the process; i.e., the process becomes faster for larger $\widehat{K}$ without any other changes. Moreover, the system can be given a variational form as $\frac{d\theta_i}{dt} = -\widehat{K} \frac{\partial}{\partial \theta_i} G$, where $G = -\frac{1}{2}\sum_{i,j=\langle i \rangle} \cos(\theta_j - \theta_i)$. Thus, the system evolves with time toward a minimum of $G$. The global minimum of $G$ corresponds to $\theta_j = \theta_i$ for all coupled pairs, suggesting that perfect alignment should eventually be achieved.

When a random initial condition for the $\theta_i$ values is employed, many topological defects typically arise. The coherence length increases with time as the number of defects decrease by annihilation, as observed in various systems (see Ref. [37] and references therein). Thus, with a given observation time, a larger coherence length is obtained for larger $\widehat{K}$. We suppose that this speed-up effect is a primal mechanism underlying the increase of coherence length with the nematic coupling strength, denoted by $K$ in our model of motile cells. These observations are in line with Ref. [12], where the author observes pattern formation in cultures of fibroblasts with elongated cell shapes and finds that the number of swirling patches progressively decreases after confluence and a single parallel array is eventually obtained.

Correlation analysis reveals that the patterns formed under the conditions of this experiment are on the order of several cell-lengths long; however, larger cell assemblies may migrate collectively *in vivo* [38]. Possible factors that hamper larger scale coherence *in vitro* include noisy single-cell migration, cell division, and topological defects [10]. To realise the large-scale collective migration *in vivo*, several other factors may play substantial or complementary roles, such as cell adhesion, spontaneous role assignment among cells, and polarity alignments among adjacent cells [38-40]. Mechanical tension applied to the tissue would also contribute to collective migration [41, 42].

Our experimental and numerical results support our hypothesis that cell shape affects large-scale coherence by mediating the strength of cell-cell contact interactions. Although our finding is based on an *in vitro* system, such a mechanism may be at work *in vivo* as well. Thus, our study suggests that cell shape may play an essential role in cell-cell communication in single and multi-cellular organisms.

## 4. Materials and methods

**Coating of culture dish.**

Collagen type-I, collagen type-IV, and gelatin were purchased from Nitta Gelatin Inc. (Osaka, Japan. Product names: Cellmatrix Type-I-A, Cellmatrix Type-IV, and GLS250). Collagen type-I and type-IV are provided as acidic solutions of pH 3 and concentration 3 mg/mL. These were diluted with a 1 mM HCl solution (pH 3.0) to the desired concentrations. Gelatin was dissolved in the 1 mM HCl solution to the desired concentration. Maintaining the acidic condition prevents the formation of collagen fibrils from the collagen molecules. The heat-denatured collagen type-I solution was prepared by heating the 10.0 μg/ml Cellmatrix Type-I-A solution for 30 minutes at 60°C.

The polystyrene culture dishes used in this experiment were 100 mm in diameter and obtained from AS ONE (Osaka, Japan). The glass culture dishes were 35 mm in diameter and obtained from Iwaki Co., Ltd. (Tokyo, Japan).

The culture dishes were incubated with the various types of collagen solutions, or just the vehicle solution (1 mM HCl) for the uncoated control, overnight at 4°C. Subsequently, the dishes were air dried and washed four times in phosphate-buffered saline (PBS).

The amount of collagen that was attached to the dish surface was estimated as follows. Following the same procedure as the culture dishes, each well of a 96-well polystyrene plate was incubated with 50 μL of the collagen or vehicle solution overnight at 4°C, then air dried and washed four times in PBS. The amount of collagen attached to the well surface was then measured with the Collagen Quantitation Kit (Cosmo Bio, Tokyo, Japan) following the manufacturer's protocol. By comparing the amount of coated collagen prepared with 1 and 10 μg/mL collagen solutions, we confirmed that the collagen coated on the wells increased approximately in proportion to the dosage (Fig. S1). Note that the detection limit of the Quantitation Kit was 0.4 μg/mL, hence the amount of collagen in the well prepared with the 0.1 μg/mL solution was below the detection limit.

**Cell culture.**

Normal human-skin fibroblasts, RIKEN original (NB1RGB), were provided by the RIKEN BioResource Research Center through the National BioResource Project of MEXT, Japan. The cells were cultured in minimum essential medium alpha (MEMα; Life technologies, Carlsbad, CA) supplemented with 10% fetal bovine serum (FBS; Biowest, Nuaille, France) at 37°C in a humidified incubator with a 5% $CO_2$ atmosphere.

The 100-mm diameter polystyrene dishes were seeded with $5.0 \times 10^5$ NB1RGB cells, and incubated for up to 144h (6 days). The exception was the experiment reported in Fig. S2, which started out with $1.0 \times 10^6$ cells and incubated for 72 hours (3 days), the larger initial cell-count leading to an earlier attainment of confluence. The 35-mm diameter glass dishes were seeded with $6.0 \times 10^4$ NB1RGB cells so that the initial cell-density will be the same as the 100-mm diameter dish with $5.0 \times 10^5$ cells. These were incubated for up to 90h. After incubation, the dishes were washed in ice-cold PBS and the cells fixed in ice-cold 100% methanol.

**Cell number measurement.**

For the cell-number density measurements reported in Figs. 2B and S3C, the cells were dissociated from the dishes with 0.025% Trypsin-EDTA, and the number of cells counted using an automated cell counter (BACKMAN COULTER, Brea, CA).

**Orientation Analysis.**

For cell-orientation analyses reported in Figs. 1, S2, and S3, images of the swirling patterns were captured at ×50 magnification with a digital microscope (VH-Z20W, KEYENCE, Osaka, Japan). Each image had 1600 × 1200 pixels, corresponding to an area of 6960 × 5220 µm². Since this is quite small compared to the total area of the 100-mm diameter polystyrene dish, images of 10 different non-overlapping fields of the dish were collected from each. For the 35-mm diameter glass dishes, with a 12% area compared to the 100-mm dish, one image was taken from each. These images were analysed with the ImageJ plugin OrientationJ [19] to generate the colour-coded maps shown in Figs. 1A, S2A-E, and S3A.

OrientationJ determines the local orientation $\theta$ of an image as follows. The 2D monochrome image is essentially an intensity function $f(x, y)$ defined for every pixel $(x, y)$ of the frame. OrientationJ overlays a *Gaussian window* $w(x - x_0, y - y_0)$ on the field and computes the *structure tensor* matrix

$$J_{ij}(x_0, y_0) = \int w(x - x_0, y - y_0) \, \partial_i f(x, y) \, \partial_j f(x, y) \, dx \, dy$$

for every $(x_0, y_0)$. Here, $w(x - x_0, y - y_0)$ is a gaussian centered at $(x_0, y_0)$ with user specified width, but with its tail truncated outside the local region of interest. The *dominant orientation* $\vec{u} = (\cos \theta, \sin \theta)$ at $(x_0, y_0)$ is a vector of norm 1 which maximizes

$$u_i \, J_{ij}(x_0, y_0) \, u_j = \|\partial_\theta f\|^2$$

i.e. the norm of the directional derivative of $f(x_0, y_0)$ in the direction of $\vec{u}$. It is the normalized eigenvector of the largest eigenvalue of the structure tensor matrix $J_{ij}(x_0, y_0)$. The value of the orientation $\theta$ is then colour-coded according to the scale shown on the right-margin of Fig. 1B.

To obtain the average-orientation images, each 1600 × 1200-pixel image was subdivided into a 50 × 38 grid, each subdivision being 32 × 32 pixels (139.2 × 139.2 µm²) in size. We label the subdivisions with a pair of integers $(k, \ell)$, where $1 \leq k \leq l_x = 50$, and $1 \leq \ell \leq l_y = 38$. The index $k$ labels the columns of the grid from left to right, while $\ell$ labels the rows of the grid from top to bottom. The orientation $\theta_{k\ell}$ of the subdivision $(k, \ell)$ was then obtained by setting a Gaussian window of the size of 32 × 32 pixels in OrientationJ.

**Correlation functions of average orientation.**

The correlation functions of average orientation are computed as follows. The correlation between region $(i, j)$ and region $(k, \ell)$ is defined as $C_{i,j,k,\ell} = \cos 2(\theta_{ij} - \theta_{k\ell})$. The total correlation between region $(i, j)$ and all other regions at distance $d$ from region $(i, j)$ is calculated as follows:

$$C_{i,j}(d) = \sum_{(i-k)^2 + (j-\ell)^2 = d^2} C_{i,j,k,\ell}$$

The correlation function $C(d)$ is computed as the average of this total correlation over all regions:

$$C(d) = \frac{1}{l_x l_y} \sum_{i=0}^{l_x - 1} \sum_{j=0}^{l_y - 1} C_{i,j}(d)$$

The distance value $d_{50}$ is defined as $C(d_{50}) = 0.5$. Note that the distance $d$ in these expressions is given in units of 32 pixels (139.2 µm) so it must be multiplied by 139.2 µm to convert to physical units.

**Cell movies.**

To obtain high-resolution images, we used the 35-mm diameter glass-bottom dishes

(Iwaki Co., Ltd., Tokyo, Japan) along with a high numerical aperture objective lens. The smaller size of the glass dishes allowed us to place up to three dishes simultaneously inside a temperature- and humidity-controlled microscope (BZ-X700, KEYENCE, Osaka, Japan), enabling continued observation of multiple cell cultures incubating under identical conditions.

The glass-bottom culture dishes were collagen-coated following the procedure described above from the 10.0 μg/mL collagen type-I solution. Uncoated controls were prepared using only the vehicle solution (1 mM HCl) in the same procedure. NB1RGB cells ($6.0 \times 10^4$ cells/35-mm diameter dish) were seeded in the glass-bottom dishes and cultured for 90 h while taking time-laps images at 15-min intervals using a microscope (BZ-X700, KEYENCE, Osaka, Japan). See Supplementary Videos 1 and 2.

The VW-H2MA motion analyser (KEYENCE), which performs cell tracking, was used to measure the cell migration velocity. The displacement of each isolated cell during each 15-minute interval was measured, from which the average velocity of each cell during that time-interval (which is essentially the instantaneous velocity due to the cells moving slowly) was determined (Fig. 2A). This velocity was averaged over twenty cells. The instantaneous velocities for low- and high-density conditions were respectively based on data from 0–30 h and 60–90 h after culture start. The 10h-average velocity, shown in Fig. S3D, is based on the linear displacement during each 10 h interval. This velocity was averaged over thirty cells. The 10-h average velocities for low- and high-density conditions were respectively based on data from 0–30 h and 60–90 h after culture start (Fig. S3E). The area $S$ and perimeter $L$ of the NB1RGB cells were measured using ImageJ after culturing for 24 h, and circularity was calculated as $4\pi S/L^2$ (Figs. 2E and 2F).

**Mathematical model.**

To obtain insight on the role of collagen in the 2D patterns formed by fibroblasts, we introduced a simple mathematical model of cell collective motion. The model equation is given as:

$$\frac{dx_i}{dt} = v \cos \theta_i + R_x(i) \tag{1}$$

$$\frac{dy_i}{dt} = v \sin \theta_i + R_y(i) \tag{2}$$

$$\frac{d\theta_i}{dt} = K \sum_j \exp\left(-\frac{r_{ij}^2}{2\lambda^2}\right) \sin 2(\theta_j - \theta_i) + \sqrt{\mu}\xi_i \tag{3}$$

where the variables $(x_i(t); y_i(t))$ and $\theta_i(t)$ are the position and the orientation of cell $i$ at time $t$, respectively. In this model, cell $i$ migrates spontaneously in the direction $\theta_i(t)$ with constant speed $v$. The terms $R_x$ and $R_y$ denote the repulsive force due to cell–cell excluded-volume interactions described as a Gaussian soft-core potential $H(r) = \sigma^2 K_C \exp\left(-\frac{r^2}{2\sigma^2}\right)$ with interaction strength $K_c$, distance $r$ between cells, and repulsion length $\sigma$. Explicitly, the terms $R_x$ and $R_y$ are given as:

$$R_x(i) = -\frac{\partial}{\partial x_i} \sum_j H(r_{ij}) = K_c \sum_j (x_i - x_j) \exp\left(-\frac{r_{ij}^2}{2\sigma^2}\right)$$

$$R_y(i) = -\frac{\partial}{\partial y_i} \sum_j H(r_{ij}) = K_c \sum_j (y_i - y_j) \exp\left(-\frac{r_{ij}^2}{2\sigma^2}\right)$$

where $r_{ij}$ stands for the distance between cell $i$ and cell $j$. Equation (3) describes the nematic interaction with strength $K$ between the cells that leads to the nematic alignment of cell orientations. A similar interaction was considered by Sumino Y. *et al.*[7]. The effective interaction strength in Equation (3) is $K \exp\left(-\frac{r_{ij}^2}{2\lambda^2}\right)$ with characteristic interaction length $\lambda$. An additive noise is also introduced in Eq. (3), where $\mu$ is the noise strength and $\xi_i(t)$ is white noise with zero mean and unit variance, i.e., $\langle \xi_i(t)\xi_j(s)\rangle = \delta_{ij}\delta(t-s)$.

**Fixed parameters.**

The interaction lengths $\sigma$ and $\lambda$ are both fixed to $\sigma = \lambda = 0.696$ units (70 μm) in all simulations. We take 140 μm to be the typical cell length, cf. Fig.2C and D, and have chosen $\sigma$ and $\lambda$ to be ½ of this value. Except when the excluded volume effect is turned off for Fig.4, case (vi) ($K_C = 0$), the soft-core interaction strength is taken to be $K_C = 0.89$/h. This value was found via trial and error to reproduce the experimental characteristics well.

**Simulations.**

The continuous-time model of Equations (1)–(3) was discretised for each cell using Euler's method with a step size of 30 min, for a simulation length of 144 h (6 days), resulting in $144 \times 2 + 1 = 289$ frames.

To match the experimental data, each frame's dimensions were adjusted to 69.6 × 52.2 units, where 1 unit is 100 μm. Instead of considering periodic boundaries, which would not match the experimental conditions, we added 15-unit wide margins to the simulation frames. Therefore, the actual frame dimensions for the simulation were 99.6 × 82.2 units, within which cells were initially placed. When creating the simulation video and while taking quantitative measurements for comparison with the experimental results, these margins were ignored.

The initial number of cells is $N_0 = 3434$, which corresponds to the initial cell density in the experiments, ~ 8000 cells/cm². All the simulated cells were initialised with random positions $(x_i(0), y_i(0))$ in the window and with a random orientation $\theta_i(0)$ in the range $[0, 2\pi)$.

Based on Fig. 2B, we assume that the increase in the cell number induced by cell divisions starts at $t = 24$ h and is set to 1.005 cells/0.5h so that approximately 11,000 cells result after 144 h of cultivation. Each new cell is divided from a randomly chosen existing cell, and positioned 0.35 units (35 μm) away from the parent cell, either in front or behind on the parent's line of motion with equal probability, and with the same orientation and direction of motion. The distance of 35 μm ($= \sigma/2$) was chosen so that the position of the offspring will initially lie within the length of the parent, cf. Fig.2C and D. We do not take into account cell cycles; i.e., each cell proliferates irrespective of its proliferation history. We expect that the detail of the proliferation rule does not considerably affect our results since the cell number is large enough for random and cyclic proliferation rules to be statistically almost equivalent.

GIF animations from the 289 frames were generated using Gnuplot.

The block-average orientation $\theta_{k\ell}$ of the cells in region $(k, \ell)$ is defined as the exponent appearing in the equation below:

$$R_{k\ell} e^{i2\theta_{k\ell}} = \frac{1}{N_{k\ell}} \sum_{j=0}^{N_{k\ell}-1} e^{i2\theta_j}$$

where $i$ is the square root of $-1$, and $N_{k\ell}$ is the number of cells in region $(k, \ell)$. Correlation data were then measured at $t = 6, 24, 48, 72, 96, 120$, and 144 h, following the protocol described above for the experimental observations.

To correct for the effects of randomness in the simulation, 10 simulations with the same parameters were run and averaged data were reported as the simulation results.

**Dependence of coherence length on migration speed.**

The simulations were conducted for the following three cases with regards to the constant speed $v$:

1) The measured migration velocity at low density on an uncoated dish: $v = 0.1864$ unit/h (1 unit $= 100$ μm, cf. Fig. 2A),

2) The measured migration velocity at low density on a coated dish: $v = 0.1598$ unit/h (cf. Fig. 2A),

3) The average of the above two: $v = 0.1731$ unit/h.

The dependence of the coherence length on these selections is shown in Fig. 3G. The solid, dashed, and dotted lines are the average $d_{50}$ values over ten simulations for cases 1), 2) and 3), respectively. The error bars are the standard deviation of 10 runs for case 1).

**Statistical analyses.**

The data were analysed with the one-tailed Student's *t*-test. The values were expressed as the mean ± mean standard error. Changes were considered to be significant if the *p* value from the Student's *t*-test was less than 0.05.

## Acknowledgements

This work was supported by grants from the Leading Graduate School Promotion Center, Ochanomizu University. We are grateful to Dr. Khayrul Bashar, Dr. Kyogo Kawaguchi and Dr. Kyohei Shitara for their helpful advice. We are also grateful to Dr. Daiki Nishiguchi for providing the ImageJ script for automatic image analysis.

K.Y. acknowledges support from the Leading Graduate School Promotion Center of Ochanomizu University for a long-term stay at Virginia Tech; she thanks Professor John J. Tyson and his Computational Cell Biology Lab at Virginia Tech for their warm hospitality and the valuable advice she received while this work was conducted. M.G. acknowledges the financial support from JSPS KAKENHI Grant No. 26860144. H.K. acknowledges the financial support from MEXT KAKENHI Grant No. 15H05876 and JSPS KAKENHI Grant No. 18K11464.

## Compliance with ethical standards

**Conflict of interest** The authors declare that they have no conflict of interest.

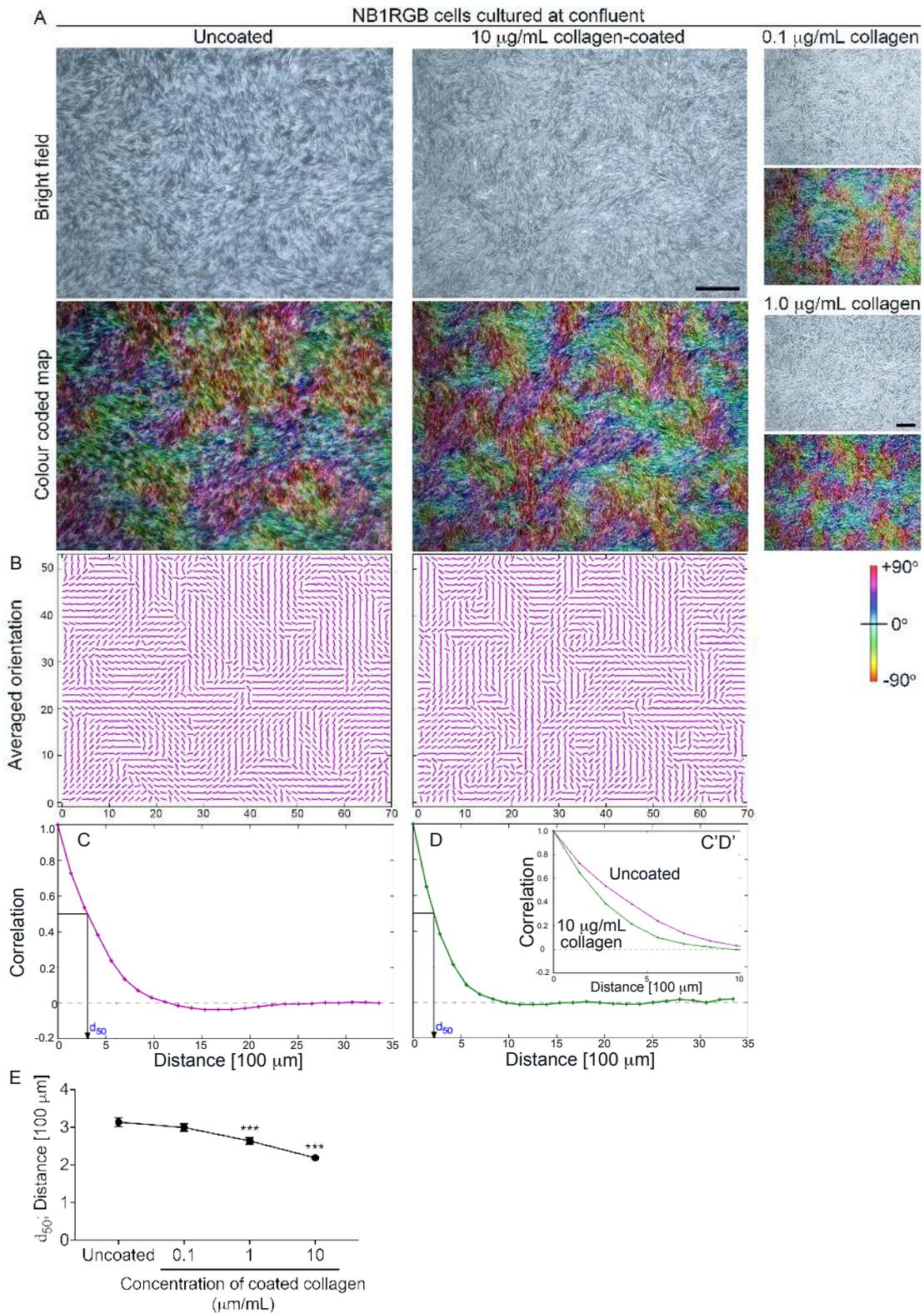

**Fig.1.** Patterns formed by human-skin NB1RGB fibroblast cells cultured in a polystyrene dish.

(A) Typical patterns of confluent fibroblasts cultured for 144 h on the uncoated control (left) and the dish coated from the 10.0 µg/mL collagen type-I solution (right). The black scale bar in the lower-right corner of the right-monochrome image is 1 mm long. The smaller images in the right-margin are dishes coated from 0.1 µg/mL (top) and 1.0 µg/mL (bottom) collagen type-I solutions. For each pair of images, the lower colour-map encodes the orientations of the fibroblasts in the upper bright-field image in accordance with the colour-scale shown to the right of B.

(B) The block-averaged orientations of the fibroblasts on uncoated (left) and 10.0 µg/mL-collagen-coated (right) dishes shown in A.

(C, D) Correlation functions of cell orientation for the uncoated (C), and collagen-coated (D) cases. The graphs C′ and D′ in the sub-window show the behaviour of the two functions near the origin on the same axes for the ease of comparison.

(E) $d_{50}$ values averaged over 40 images. Data represent the mean ± standard error of mean (SEM) from 4 independent cultures. Ten images were captured from each culture dish. Datapoints labelled with a triple asterisk (***) indicate points for which $p < 0.001$ under Student's one-tailed *t*-test when compared to the control.

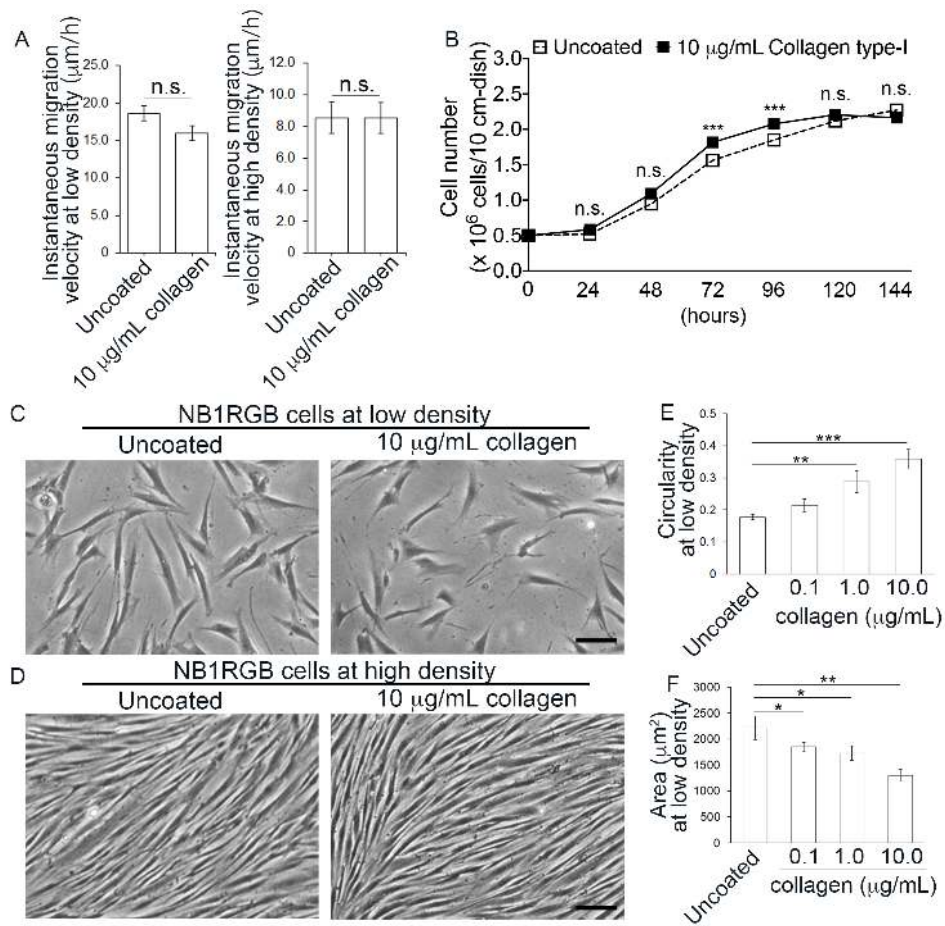

**Fig. 2.** Effect of collagen coating on the motility (A), number (B), and morphology (E,F) of human-skin fibroblast cells. Samples are compared with Student's *t*-test and labeled n.s. (no significance), * ($p < 0.05$), ** ($p < 0.01$), or *** ($p < 0.001$).

(A) Migration velocities of the fibroblasts cultured at low and high densities on uncoated and 10.0 μg/mL-collagen-type-I-coated glass dishes. Each data point represents the mean ± SEM of 20 cells chosen randomly from two cell cultures.

(B) Number of fibroblasts cultured for 24-144 h on uncoated and 10 mg/mL-collagen-type-I-coated polystyrene dishes. Data points represent the mean ± SEM of 4 dishes.

(C through F) Analysis of fibroblast cell morphology.

(C, D) Fibroblasts cultured on uncoated and 10 μg/mL-collagen-type-I-coated polystyrene dishes at 24 h (C; low density) and 72 h after culture start (D; high density).

Scale bar: 100 μm. The fibroblasts were observed and photographed with a Nikon ECLIPSE TS 100 phase-contrast microscope (NIKON Corp., Tokyo, Japan).

(E, F) Circularity and area of fibroblasts cultured on uncoated and 0.1, 1.0, and 10.0 μg/mL collagen type-I coated polystyrene dishes at low density. Each data point represents the mean ± SEM of 20 cells chosen randomly from two cell cultures.

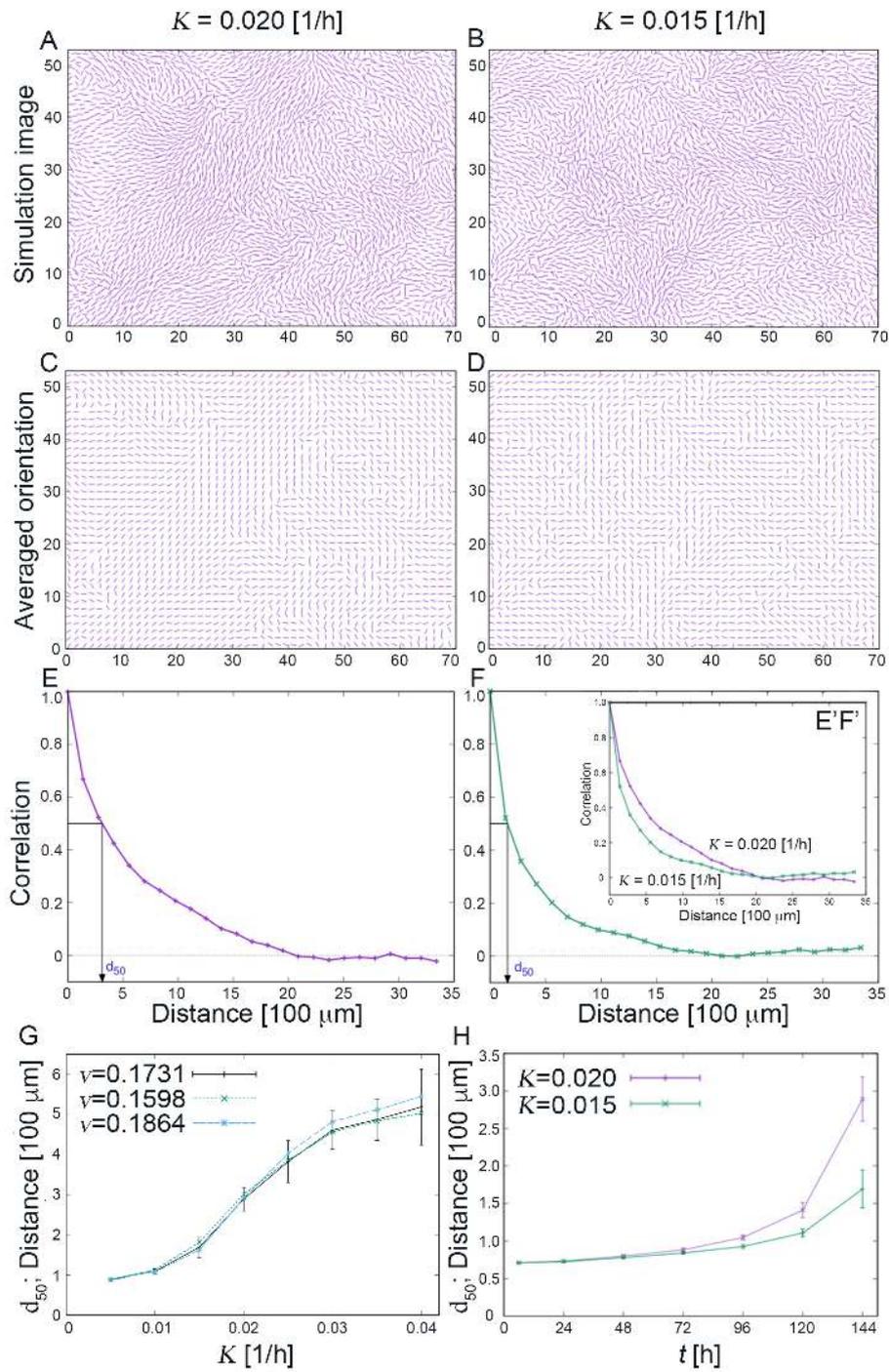

**Fig. 3.** Simulation results. The strength of the nematic interaction is $K = 0.020$/h in (A, C, E) and $K = 0.015$/h in (B, D, F).

(A, B) Cell positions and orientations.

(C, D) Averaged orientations for the 32 × 32 pixel subdivisions.

(E, F) Typical correlation functions of the averaged cell orientations and the determination of $d_{50}$. The subwindow labelled E'F' in F shows the same graphs together on the same axes for the ease of comparison.

(G) Dependence of $d_{50}$ on the interaction strength $K$ for three different choices of the cell migration speed $v$. Dashed line: $v = 0.1864$ unit/h (uncoated measured value), dotted line: $v = 0.1598$ unit/h (10 µg/mL-collagen-type-I-coated measured value), solid line: $v = 0.1731$ unit/h (average of the two measurements). The values of $d_{50}$ are averaged over 10 simulations for each value of $K$. For the $v = 0.1731$ unit/h (average) case, error bars are shown for each point.

(H) Time courses of $d_{50}$ for $K = 0.020$/h and $K = 0.015$/h.

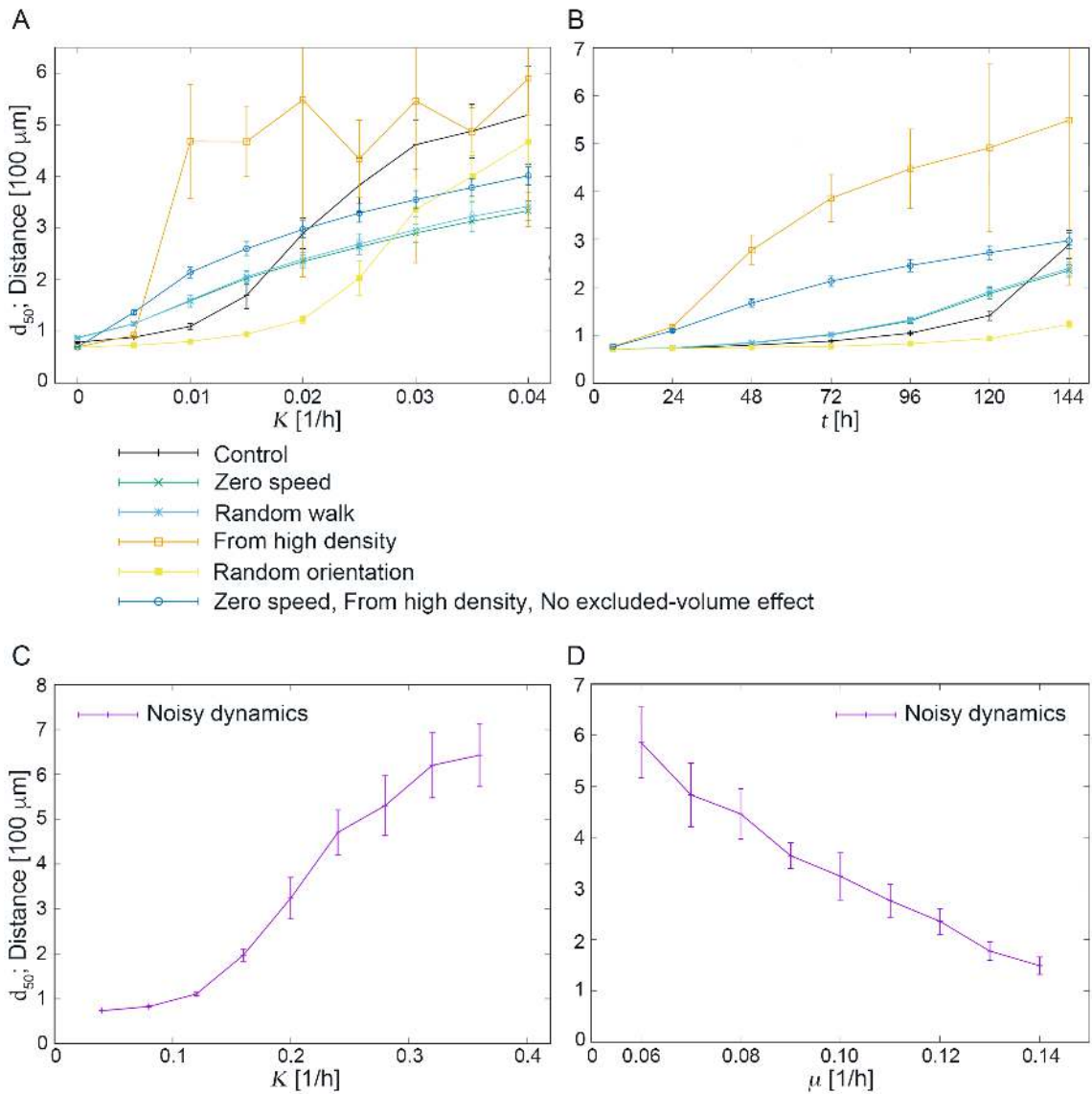

**Fig. 4.** Simulations results.

(A) $d_{50}$ vs $K$ at $t$ =144h and

(B) $d_{50}$ vs $t$ for $K$ =0.02/h with the following conditions:

(i) "control", which is the same as the result shown in Fig. 3 with $v$ = 0.1864 units/h,

(ii) "zero speed", where $v$ = 0,

(iii) "random walk", where the migration direction of each cell at each time step is set to be a random value irrespective of its orientation,

(iv) "from high density", where the cell proliferation is absent and the initial number of cells is 14,080,

(v) "random orientation", where the orientation of the newborn cell is assigned a random value,

(vi) "zero speed, from high density, no excluded volume effect", where we employ (ii) and (iv) and further assume $\sigma = 0$.

(C) $d_{50}$ vs $K$ at $t = 144$h with noise strength $\mu = 0.1$/h.

(D) $d_{50}$ vs $\mu$ at $t = 144$h for $K = 0.2$/h .

SUPPLEMENTARY INFORMATION

# The effects of coating culture dishes with collagen on fibroblast cell shape and swirling pattern formation

Kei Hashimoto[1,2,3,†], Kimiko Yamashita[1,2,4,5,†], Kanako Enoyoshi[1,2,†], Xavier Dahan[2], Tatsu Takeuchi[6], Hiroshi Kori[1,7,*], and Mari Gotoh[3]

[1]Graduate School of Humanities and Sciences, Ochanomizu University, Ohtsuka, Bunkyo-ku, Tokyo, Japan

[2]Program for Leading Graduate Schools, Ochanomizu University, Ohtsuka, Bunkyo-ku, Tokyo, Japan

[3]Institute for Human Life Innovation, Ochanomizu University, Ohtsuka, Bunkyo-ku, Tokyo, Japan

[4]Department of Physics, National Tsing Hua University, Hsinchu, Taiwan

[5]Physics Division, National Center for Theoretical Sciences, Hsinchu, Taiwan

[6]Department of Physics, Virginia Tech, Blacksburg, VA 24061, USA

[7]Department of Complexity Science and Engineering, Graduate School of Frontier Sciences, The University of Toyo, Kashiwa, Japan

*Correspondence should be addressed to H. Kori (kori@k.u-tokyo.ac.jp)

**Supplementary Figure S1**

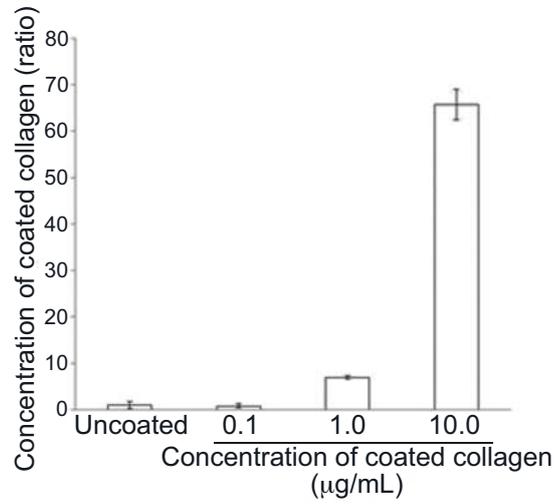

The concentration of coated collagen on the well surface. The ratio of coated collagen type-I concentration on the well after incubating with 0 (uncoated control, only the 1 mM HCl vehicle solution), 0.1, 1.0 and 10.0 µg/mL collagen type-I solutions. The concentrations are normalized to that of the uncoated well. Data represent the mean ± SEM for each of the four initial collagen concentrations.

**Supplementary Figure S2**

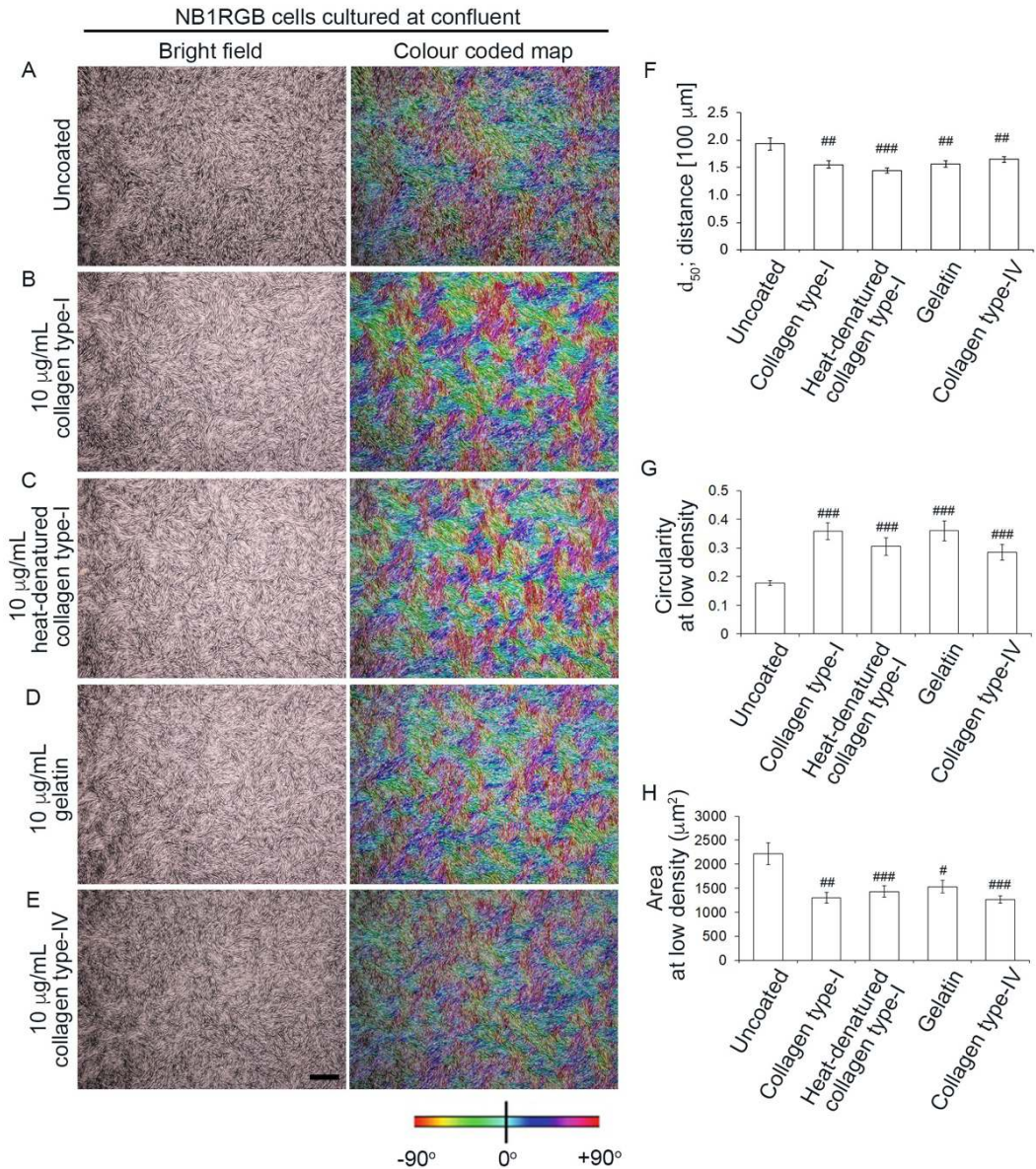

The effects of various types of collagen on the pattern formation of human-skin fibroblasts, NB1RGB, cultured for 72h on polystyrene dishes. (A–E) Bright-field images and colour coded maps of cultured NB1RGB cells at confluence on uncoated (control) (A), 10.0 μg/mL collagen type-I-coated (B), 10.0 μg/mL heat-denatured collagen type-I-coated (C), 10.0 μg/mL gelatin-coated (D), and 10.0 μg/mL collagen type-IV-coated (E) dishes. Scale bar: 1 mm. (F) $d_{50}$ values at 72 h. #: $p < 0.05$, ##: $p < 0.01$, ###: $p < 0.001$, $t$-test vs. the $d_{50}$ value of control. (G, H) Circularity and area of NB1RGB cultured at 24 h. #: $p < 0.05$, ##: $p < 0.01$, ###: $p < 0.001$, $t$-test vs. control value. Correlation data represent the mean ± SEM of 12 images from 4 dishes. Other data represent the mean ± SEM of 20 cells.

**Supplementary Figure S3**

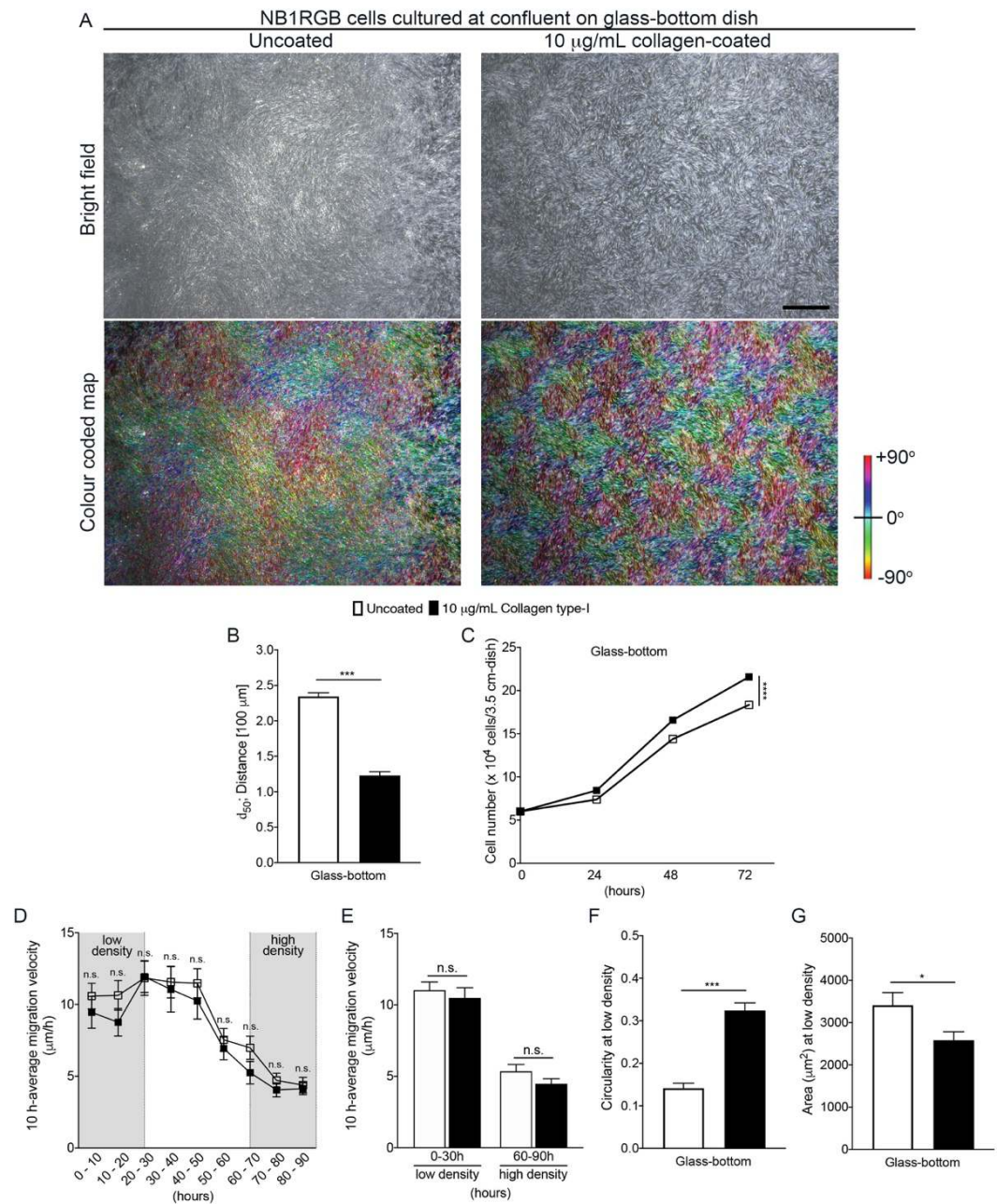

Patterns formed by human-skin NB1RGB fibroblast cells cultured on glass-bottom dishes. (A) Bright-field images and colour coded maps of NB1RGB cultured for 72 h at confluence on uncoated and 10.0 μg/mL collagen type-I-coated glass-bottom dishes. Scale bar: 1 mm. (B) $d_{50}$ value of NB1RGB cells cultured for 72 h. ***: $p < 0.001$, $t$-test. Data represent the mean ± SEM of 4 images from 4 dishes. (C) Number of cells on glass-bottom dishes. ****: $p < 0.0001$, two-way ANOVA. Data represent the mean ± SEM of 4 dishes. (D, E) 10 h-average migration velocity of NB1RGB cultured for 90 h on glass-bottom dishes. two-way ANOVA. Data represent the mean ± SEM of 40 cells at least. (F, G) Circularity and area of NB1RGB morphology at 24 h. *: $p < 0.05$, ***: $p < 0.001$, *t*-test. Data represent the mean ± SEM of 30 cells.

**Supplementary video legends**

(Not available in this preprint).

**Video 1** Experimentally obtained time courses of cell growth and migration on the uncoated glass-bottom dish.

**Video 2** Experimentally obtained time courses of cell growth and migration on the 10 μg/mL collagen type-I-coated glass-bottom dish.

**Video 3** Numerically obtained time courses of cell growth and migration for $K$ = 0.020/h.

**Video 4** Numerically obtained time courses of cell growth and migration for $K$ = 0.015/h.

**Video 5** Numerically obtained time courses of alignment process in the absence of spontaneous mobility, reproduction, and excluded volume effect. $K$ = 0.020/h.

**Video 6** Numerically obtained time courses of cell growth and migration for $K$ = 0.20/h and $\mu = 0.10$/h.

**Video 7** Numerically obtained time courses of cell growth and migration $K$ = 0.40/h and $\mu = 0.10$/h.